\documentclass[twocolumn]{openjournal}

\usepackage{amsmath,amssymb,mathtools,bm}
\usepackage{slashed}
\usepackage{aas_macros}
\usepackage{hyperref}

\newcommand{\U}{\hat{u}}
\newcommand{\E}{\mathrm{e}}

\newcommand{\D}{\mathrm{d}}
\newcommand{\C}{\mathrlap{\slashed{\phantom{C}}}C}
\newcommand{\B}{\mathrlap{\slashed{\phantom{B}}}B}

\DeclarePairedDelimiter{\Ev}{\langle}{\rangle}

\makeatletter
\def\@threej(#1,#2,#3,#4,#5,#6){%
\begin{pmatrix}#1&#3&#5\\#2&#4&#6\end{pmatrix}}
\newcommand{\threej}[3]{\@threej(#1,#2,#3)}
\makeatother

\begin{document}

\title{Shot noise in clustering power spectra}

\author{Nicolas Tessore}
\email{n.tessore@ucl.ac.uk}
\affiliation{Department of Physics and Astronomy, University College London, Gower Street, London, WC1E\,6BT, UK}
\affiliation{Mullard Space Science Laboratory, University College London, Holmbury St Mary, Dorking, RH5\,6NT, UK}

\author{Alex Hall}
\email{ahall@roe.ac.uk}
\affiliation{Institute for Astronomy, University of Edinburgh, Royal Observatory, Blackford Hill, Edinburgh EH9\,3HJ, UK}

\journalinfo{The Open Journal of Astrophysics}
\submitted{July 2025}

\begin{abstract}
\noindent
We show that the `shot noise' bias in angular clustering power spectra observed from discrete samples of points is not noise, but rather a known additive contribution that naturally arises due to degenerate pairs of points.
In particular, we show that the true shot noise contribution cannot have a `non-Poissonian' value, even though all point processes with non-trivial two-point statistics are non-Poissonian.
Apparent deviations from the `Poissonian' value can arise when significant correlations or anti-correlations are localised on small spatial scales.
However, such deviations always correspond to a physical difference in two-point statistics, not a difference in noise.
In the context of simulations, if clustering is treated as the tracer of a discretised underlying density field, any sub- or super-Poissonian sampling of the tracer induces such small-scale modifications and vice versa; we show this explicitly using recent innovations in angular power spectrum estimation from discrete catalogues.
Finally, we show that the full covariance of clustering power spectra can also be computed explicitly: it depends only on the two-, three- and four-point statistics of the point process.
The usual Gaussian covariance approximation appears as one term in the shot noise contribution, which is non-diagonal even for Gaussian random fields and Poisson sampling.
\end{abstract}

\maketitle

\section{Introduction}

The observed clustering of galaxies is an important probe of cosmological information, either as angular clustering on the sphere \citep[recently][]{2022PhRvD.105b3520A, 2023A&A...675A.189D, 2023PhRvD.108l3520M} or as 3D clustering in redshift space \citep[recently][]{2025arXiv250314738D}.
The two-point statistics of galaxy clustering are principally measured in two ways: as correlation functions in real (or configuration) space, or as power spectra in harmonic (or Fourier) space.
The latter are attractive for several reasons.
On a practical level, measuring power spectra is typically much faster than measuring correlation functions.
In addition, power spectra are typically less correlated between scales compared with their configuration space counterparts, and allow a cleaner separation of non-linear scales.

However, clustering power spectra measured from sets of observed points contain an obvious contamination, commonly called `shot noise'.
Shot noise is an additive contribution to the power spectra with a constant value at all scales.
This constant shot noise contribution must therefore in general be subtracted to obtain an estimate of the actual power spectra of interest.

It would appear that shot noise arises in the following manner: there is an underlying density field, which contains the two-point statistics of interest, and of which the observed points are `tracers'.
These tracers are drawn randomly from the underlying density field, following some sampling distribution, and this random sampling introduces additional variance into the measurement of the power spectra.
The additional variance is a delta function at zero lag in the correlation function, which translates to a constant additive term --- shot noise --- in the power spectrum.
In this picture, shot noise is closely related to the random sampling of the tracers.
Therefore, one might worry that a model for the clustering of the points on very small scales, or indeed the distribution of number counts in very small cells, is required in order to accurately model and subtract the shot noise contribution from measured power spectra.

In this work, we aim to clarify the distinction between `shot noise' and the nature of the point process that generates the observation.
We show that the true shot noise contribution to the power spectrum, which is constant on all scales, depends only on the observed number of points, and not on their sampling distribution.
This shot noise contribution always has the `Poissonian' value, and any apparent deviation from said value corresponds to a change in the underlying two-point statistics.
These two-point statistics describe real physical effects \citep[e.g.][]{2013PhRvD..88h3507B, 2017PhRvD..96h3528G, 2018A&A...615A...1L} that have non-noise power spectra, even if this is not apparent at the scales at which the spectra are measured.

In what follows, we establish a series of facts about point processes and shot noise.
\begin{itemize}
\item The distribution of clustered points is always non-Poissonian (Section~\ref{sec:clustering}).
\item The `shot noise' contribution to clustering power spectra is always `Poissonian' (Section~\ref{sec:shotnoise}).
\item Any apparent deviation from the Poissonian shot noise contribution comes from a difference in two-point statistics, not shot noise (Section~\ref{sec:tracers}).
\item The shot noise contribution to the covariance of observed angular power spectra can be computed exactly (Section~\ref{sec:covariance}).
\end{itemize}
A short summary of the conclusions from these facts is given in Section~\ref{sec:conclusions}.

\section{Clustering is always non-Poissonian}
\label{sec:clustering}

For a homogeneous and isotropic point process such as galaxy clustering, the joint probability distribution of a pair of points~$\U$ and~$\U'$ can be written in terms of the two-point correlation function~$w$ as \citep{1973ApJ...185..413P}
\begin{equation}
\label{eq:pdf}
    p(\U, \U')
    \propto 1 + w(\theta) \;,
\end{equation}
where the angle~$\theta = \arccos(\U \cdot \U')$ is the angular distance between~$\U$ and~$\U'$.
The joint probability distribution~\eqref{eq:pdf} fully describes the two-point statistics of a homogeneous and isotropic point process, and no further assumptions about the nature of the point process, such as being a Poisson process, are necessary.

In fact, a correlated point process with $w \not\equiv 0$ cannot be Poissonian.
This appears to contradict the usual recipe for simulations: generate a density field~$\delta$ with correlation function~$w$, then draw the number of points in a small region~$\D\U$ from a Poisson distribution with rate~$\bar{n} \, [1 + \delta(\U)] \, \D\U$, where~$\bar{n}$ is the mean density of the field.
By construction, the resulting number density is Poissonian --- conditional on~$\delta$.
Over realisations of~$\delta$, the covariance between the numbers of points~$N_1$ and~$N_2$ in a pair of non-overlapping regions~$\Omega_1$ and~$\Omega_2$ is
\begin{equation}
    \Ev{N_1 N_2}
    - \Ev{N_1} \, \Ev{N_2}
    = \bar{n}_1 \bar{n}_2
        \int\limits_{\mathclap{\Omega_1 \times \Omega_2}} w(\theta) \, \D\U_1 \, \D\U_2 \;.
\end{equation}
Complete independence would require the integral to vanish for arbitrary~$\Omega_1 \times \Omega_2$.
The point process can therefore only be Poissonian if~$w \equiv 0$ \citep{1956AJ.....61..383L}.

\section{Shot noise is always Poissonian}
\label{sec:shotnoise}

If the joint probability distribution~\eqref{eq:pdf} fully describes the two-point statistics of the point process, the distribution of the points -- Poissonian or not -- should not enter any observable two-point quantity.
In particular, the `shot noise' contribution to the observed power spectrum cannot depend on the distribution of the points.

This can be checked explicitly by computing the value of the shot noise \citep{2025A&A...694A.141E}.
First, write the observed number density~$n$ as a sum of Dirac delta functions in the observed positions~$\U_1, \U_2, \ldots$,
\begin{equation}
\label{eq:n}
    n(\U)
    = \sum_{i} \delta^{\rm D}(\U - \U_i) \;.
\end{equation}
Next, transform the observed number density~$n$ from real space to harmonic space,
\begin{equation}
\label{eq:tfm}
    n_{\ell m}
    = \int \! n(\U) \, Y_{\ell m}^*(\U) \, \D\U \;,
\end{equation}
where~$Y_{\ell m}$ are the spherical harmonics.
Applying the transform~\eqref{eq:tfm} to the number density~\eqref{eq:n} and using the defining property of the delta function yields coefficients
\begin{equation}
\label{eq:coeff}
    n_{\ell m}
    = \sum_{i} Y_{\ell m}^*(\U_i) \;.
\end{equation}
In other words, the transform of a discrete set of points are the sums of function values of~$Y_{\ell m}$ in the observed points.

Finally, consider the observed two-point statistics in harmonic space.
Given two not necessarily distinct samples~$A$ and~$B$ of points, the observed angular power spectrum~$C_\ell^{AB}$ is
\begin{equation}
\label{eq:cl}
    C_\ell^{AB}
    = \frac{1}{2\ell + 1} \sum_{m} n_{\ell m}^{A*} n_{\ell m}^B \;.
\end{equation}
Importantly, the angular power spectrum~\eqref{eq:cl} has not been defined in a stochastic sense here, but for a given realisation of spherical harmonic coefficients.

The power spectrum~\eqref{eq:cl} for the specific coefficients~\eqref{eq:coeff} reduces to a sum over all pairs~$i, i'$ of points of the form
\begin{equation}
\label{eq:cl2}
    C_\ell^{AB}
    = \frac{1}{4\pi} \sum_{ii'} P_\ell(\cos\theta_{ii'}) \;,
\end{equation}
where~$P_\ell$ is the Legendre polynomial.
For~$\ell > 0$, the sum in~\eqref{eq:cl2} can be split into contributions from true pairs of distinct objects ($i \not\equiv i'$) and degenerate pairs of identical objects ($i \equiv i'$),
\begin{equation}
\label{eq:split}
    C_\ell^{AB}
    = \frac{1}{4\pi} \sum_{i \not\equiv i'} P_\ell(\cos\theta_{ii'})
    + \frac{1}{4\pi} \sum_{i \equiv i'} P_\ell(\cos 0) \;.
\end{equation}
Since $P_\ell(\cos 0) = 1$, the second term reduces to the number~$N_{AB}$ of degenerate pairs in samples~$A$ and~$B$,
\begin{equation}
\label{eq:nb}
    C_\ell^{AB}
    = \frac{1}{4\pi} \sum_{i \not\equiv i'} P_\ell(\cos\theta_{ii'})
    + \frac{N_{AB}}{4\pi} \;.
\end{equation}
The observed power spectrum therefore has two distinct contributions: the actual two-point statistics are encoded in a sum over true pairs of distinct objects, denoted here as the power spectrum with a slash,
\begin{equation}
\label{eq:clred}
    \C_\ell^{AB}
    \equiv \frac{1}{4\pi} \sum_{i \not\equiv i'} P_\ell(\cos\theta_{ii'}) \;,
\end{equation}
while the shot noise contribution~$N_{AB}$ is the number of objects that appear in both observations.
An exception is the monopole~$\ell = 0$: since $P_0(\cos\theta) = 1$, the value of~$\C_0^{AB}$ is always
\begin{equation}
\label{eq:c0}
    \C_0^{AB}
    = \frac{N_A \, N_B - N_{AB}}{4\pi} \;,
\end{equation}
with~$N_A$ and~$N_B$ the respective sample sizes.
As could be expected, it follows that the monopole of the power spectrum contains no information about the underlying two-point statistics.

Naturally, the split into two-point information and shot noise continues to hold when taking the expectation of the power spectrum over realisations of the observations,
\begin{equation}
    \Ev{C_\ell^{AB}}
    = \Ev{\C_\ell^{AB}}
    + \frac{\Ev{N_{AB}}}{4\pi} \;.
\end{equation}
Since~$\Ev{N_{AB}}$ is the expected number of degenerate pairs in both samples, the expected shot noise contribution to the power spectrum is independent of the distribution of the points.
And it is clear from the above that we do not require the expectation~$\Ev{N_{AB}}$ to compute the power spectra~$\C_\ell^{AB}$ with the shot noise value subtracted.
In summary, the shot noise contribution to the power spectrum does not depend on the particular process that generates the points, and always takes the `Poissonian' value.

\section{Points as tracers}
\label{sec:tracers}

Once again, it seems easy to contradict the claim of the previous section with simulations: given an arbitrary grid, one can choose to assign number counts to grid cells using an explicitly non-Poissonian distribution, which will yield an apparently non-Poissonian shot noise value.
To illustrate this, we assume a fixed ratio of variance~$\sigma_i^2$ to mean number count~$\bar{N}_i$ in each grid cell~$i$,
\begin{equation}
\label{eq:f}
    f = \frac{\sigma_i^2}{\bar{N}_i} \;.
\end{equation}
The value of~$f$ determines whether the distribution of number counts~$N_i$ is Poissonian ($f = 1$), super-Poissonian ($f > 1$), or sub-Poissonian ($f < 1$).
The mean number count~$\bar{N}_i = \bar{N} \, (1 + \delta_i)$ is conditional on the density field~$\delta_i$, where~$\bar{N}$ is the overall mean number count.
Over realisations of~$\delta$, the expected number density fluctuations are
\begin{equation}
\label{eq:var}
    \Ev[\Big]{\Bigl(\frac{N_i - \bar{N}}{\bar{N}}\Bigr)^{\!2}}
    = \Ev*{\delta_i^2} + \frac{f}{\bar{N}} \;,
\end{equation}
where, crucially, it is~$\bar{N}$ and not~$\bar{N}_i$ that normalises the density contrast in each cell.
The variance~\eqref{eq:var} shows that the shot noise bias will appear to differ from the Poissonian value by a factor of~$f$.

This apparent contradiction is resolved if we note that, in cosmological terms, the left-hand side of the variance~\eqref{eq:var} is the expected number density fluctuation~$\sigma^2(R)$ at the grid scale~$R$.
If a non-Poissonian distribution with~$f \ne 1$ changes~$\sigma^2(R)$, it must also change the number of pairs of galaxies below the grid scale.
In other words, the non-Poissonian sampling of number counts in each grid cell implicitly changes the clustering of points below the grid scale, even if those points have not been explicitly simulated.
In the observed power spectrum~\eqref{eq:nb}, the modified variance due to~$f \ne 1$ is cosmological signal coming from the sum over galaxy pairs, and must be modelled as such.
The additive contribution~$N_{AB} / (4\pi)$ always remains fixed to the Poissonian value.

This is readily demonstrated with simulations.
First, we generate a fixed realisation of a density field~$\delta$ from a theory power spectrum.
For simplicity, this is done here for an arbitrary power-law angular power spectrum.
Next, we sample the number counts~$N_i$ for tracers of this field from three different distributions:
\begin{itemize}
    \item a negative binomial distribution, which is super-Poissonian, with~$f = 2$ \citep{2025A&A...698A.148P},
    \item a Poisson distribution with $f = 1$, and
    \item a binomial distribution, which is sub-Poissonian, with~$f = 0.5$.
\end{itemize}
Each distribution has the same mean~$\bar{N}_i = \bar{N} \, (1 + \delta_i)$ in each grid cell~$i$.
Finally, we assemble a catalogue for each distribution by randomly drawing a number~$N_i$ of points from within each grid cell~$i$.

\begin{figure}[t!]%
\centering%
\includegraphics[scale=0.8]{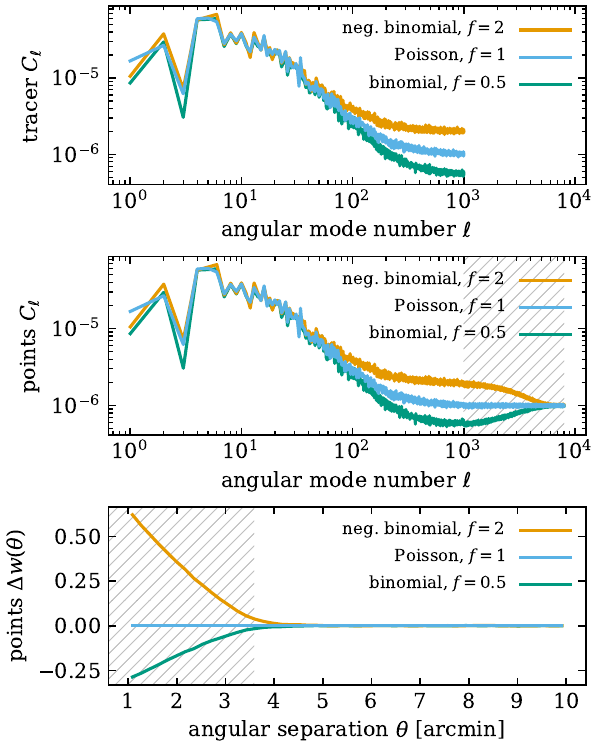}%
\caption{%
    Two-point statistics of tracers of a fixed density~$\delta$ from a negative binomial distribution, a Poisson distribution, and a binomial distribution.
    \emph{Top panel:} The angular power spectra~$C_\ell$ of the simulated maps show the apparently different `shot noise' in each realisation.
    \emph{Middle panel:} The angular power spectra~$C_\ell$ measured from the sampled catalogues reveal that the bias returns to the `Poissonian' value beyond the simulated scales (\emph{hatched}).
    \emph{Bottom panel:} The angular correlations show the corresponding differences~$\Delta w(\theta)$ in clustering with respect to the Poissonian case below the pixel scale (\emph{hatched}).
}%
\label{fig}%
\end{figure}

The resulting two-point statistics are shown in Fig.~\ref{fig}.
The angular power spectra of the simulated maps of tracer number counts show the predicted effect on the shot noise, which appears to differ from the Poissonian value by a factor of~$f$.
However, measuring the respective angular power spectra below the simulation grid scale \citep[e.g.\ directly via the coefficients~\eqref{eq:coeff}, see][]{2024JCAP...05..010B,2025JCAP...01..028W,2025A&A...694A.141E} reveals that the shot noise contribution does eventually return to the Poissonian value.
And the associated angular correlation functions show that the apparent difference in shot noise indeed corresponds to a difference in correlation below the grid scale.

\section{Covariance}
\label{sec:covariance}

Above, we have seen that the shot noise contribution to the angular power spectrum~\eqref{eq:nb} has a deterministic value that is independent of the nature of the point process.
We now show that the same also holds for the covariance of the angular power spectrum.
For special cases of shot noise, similar derivations were presented by \citet{1999MNRAS.308.1179M} for 3D clustering and by \citet{2006NewA...11..226C} and \citet{2018A&A...615A...1L} for angular clustering.
We will see that once the total number of points and the two-, three-, and four-point statistics are all fixed, the covariance matrix of the power spectra does not depend on any further properties of the point process, so the results presented in these papers generalise to an arbitrary (non-Poissonian) point process.

Beyond the monopole, which always has value~\eqref{eq:c0}, the product of two angular power spectra~\eqref{eq:clred} with the shot noise contribution subtracted is
\begin{equation}
\label{eq:clcl}
    \C_{\ell>0}^{AB} \, \C_{\ell'>0}^{A'B'}
    = \frac{1}{(4\pi)^2} \sum_{1\not\equiv2} \sum_{3\not\equiv4} P_\ell(\cos\theta_{12}) \, P_{\ell'}(\cos\theta_{34}) \;.
\end{equation}
For the sake of brevity, the numbers $1, 2, 3, 4$ refer here to the first, second, third, and fourth point in its respective sample $A$, $B$, $A'$, and $B'$.
The sums in the product~\eqref{eq:clcl} can be separated into four-point, three-point, and two-point terms,
\begin{align}
\label{eq:terms}
    \sum_{1\not\equiv2} \sum_{3\not\equiv4}
    &= \sum_{\substack{1\not\equiv2\not\equiv3\not\equiv4\not\equiv1\\1\not\equiv3\\2\not\equiv4}} \nonumber\\
    &+ \sum_{\substack{1\not\equiv2\not\equiv4\not\equiv1\\1\equiv3}}
    + \sum_{\substack{1\not\equiv2\not\equiv3\not\equiv1\\1\equiv4}}
    + \sum_{\substack{1\not\equiv2\not\equiv4\not\equiv1\\2\equiv3}}
    + \sum_{\substack{1\not\equiv2\not\equiv3\not\equiv1\\2\equiv4}} \nonumber\\
    &+ \sum_{\substack{1\not\equiv2\\1\equiv3\\2\equiv4}}
    + \sum_{\substack{1\not\equiv2\\1\equiv4\\2\equiv3}} \;.
\end{align}
This is similar to the split for the angular power spectrum~\eqref{eq:split} but at higher order: the product~\eqref{eq:clcl} contains true four-point terms plus three- and two-point terms from degenerate pairs of points.

Without loss of generality, the samples~$A, B, A', B'$ can be assumed either pairwise identical or disjoint, as it is always possible to split up intersecting samples such that this is the case.
Separating the sums~\eqref{eq:terms}, the two-point terms in product~\eqref{eq:clcl} can then be written concisely as
\begin{align}
    \C_{\ell>0}^{AB} \, \C_{\ell'>0}^{A'B'}
    &= \text{(four-point terms)} \nonumber\\
    &+ \text{(three-point terms)} \nonumber\\
    &+ \nu^{AB}_{A'B'} \, \frac{1}{(4\pi)^2} \sum_{i \not\equiv i'} P_\ell(\cos\theta_{ii'}) \, P_{\ell'}(\cos\theta_{ii'}) \;,
\label{eq:clcl:2}
\end{align}
where~$\nu^{AB}_{A'B'}$ counts the number of ways in which the sample combinations~$AB$ and~$A'B'$ agree,
\begin{equation}
    \nu^{AB}_{A'B'}
    = \delta^{\rm K}_{AA'} \, \delta^{\rm K}_{BB'}
    + \delta^{\rm K}_{AB'} \, \delta^{\rm K}_{BA'} \;,
\end{equation}
and the Kronecker symbol~$\delta^{\rm K}_{AB}$ is $1$ if samples~$A$ and~$B$ are identical, or $0$ otherwise.
We can expand the product of Legendre polynomials into a sum over single Legendre polynomials,
\begin{equation}
\label{eq:plpl:2}
    P_\ell(\cos\theta) \, P_{\ell'}(\cos\theta)
    = 4\pi \sum_{\lambda} Q_{\lambda\ell\ell'} \, P_\lambda(\cos\theta) \;,
\end{equation}
where we introduced a symbol~$Q_{\lambda\ell\ell'}$ for the coefficients involving the Wigner $3j$ symbol,
\begin{equation}
\label{eq:q}
    Q_{\lambda\ell\ell'}
    = \frac{2\lambda + 1}{4\pi} \, \threej{\lambda,0}{\ell,0}{\ell',0}^{\!2} \;.
\end{equation}
Inserting the expansion~\eqref{eq:plpl:2} into the product~\eqref{eq:clcl:2}, we recover the angular power spectrum~\eqref{eq:clred},
\begin{align}
    \C_{\ell>0}^{AB} \, \C_{\ell'>0}^{A'B'}
    &= \text{(four-point terms)} \nonumber\\
    &+ \text{(three-point terms)} \nonumber\\
    &+ \nu^{AB}_{A'B'} \sum_{\lambda} Q_{\lambda\ell\ell'} \, \C_\lambda^{AB} \;.
\end{align}
Here, we are free to write~$\C_\lambda^{AB}$ since $\nu^{AB}_{A'B'}$ vanishes unless the sample combinations $AB$ and $A'B'$ are one and the same.

The three-point contributions in the sum~\eqref{eq:terms} contain terms of the form~$P_\ell(\cos\theta_{12}) \, P_{\ell'}(\cos\theta_{13})$, corresponding to degenerate pairs with one shared point.
As above, it is possible to expand the product of Legendre polynomials,
\begin{multline}
\label{eq:plpl:3}
    P_\ell(\cos\theta_{12}) \, P_{\ell'}(\cos\theta_{13}) \\
    = (4\pi)^2 \sum_{\lambda} R_{\lambda\ell\ell'} \, Z_{\lambda\ell\ell'}(\U_1, \U_2, \U_3) \;,
\end{multline}
where now the functions~$Z_{\lambda\ell\ell'}$ form a basis for triples of points,
\begin{multline}
    Z_{\lambda\ell\ell'}(\U_1, \U_2, \U_3) \\
    = \sum_{\mathclap{\mu mm'}} \, \threej{\lambda,\mu}{\ell,m}{\ell',m'} \, Y_{\lambda\mu}^*(\U_1) \, Y_{\ell m}^*(\U_2) \, Y_{\ell'm'}^*(\U_3) \;,
\end{multline}
and we once again introduced a symbol~$R_{\lambda\ell\ell'}$ for the coefficients,
\begin{equation}
    R_{\lambda\ell\ell'}
    = \sqrt{\frac{2\lambda+1}{4\pi(2\ell+1)(2\ell'+1)}} \, \threej{\lambda,0}{\ell,0}{\ell',0} \;.
\end{equation}
Inserting the expansion~\eqref{eq:plpl:3} into the product~\eqref{eq:clcl}, we define~$\B_{\ell_1\ell_2\ell_3}^{ABC}$ as the sum over triples of distinct observed points,
\begin{equation}
    \B_{\ell_1\ell_2\ell_3}^{ABC}
    = \sum_{\mathclap{1 \not\equiv 2 \not\equiv 3 \not\equiv 1}} Z_{\ell_1\ell_2\ell_3}(\U_1^A, \U_2^B, \U_3^C) \;,
\end{equation}
which we use to write out the product~\eqref{eq:clcl} in full as
\begin{align}
\label{eq:cov}
    \nonumber
    \C_{\ell>0}^{AB} \, \C_{\ell'>0}^{A'B'}
    &= \text{(four-point terms)} \\
    \nonumber
    &+ \sum_{\lambda} R_{\lambda\ell\ell'} \, \Bigl\{
        \delta_{AA'} \B_{\lambda\ell\ell'}^{ABB'}
        + \delta_{AB'} \, \B_{\lambda\ell\ell'}^{ABA'}
        \\[-10pt]\nonumber&\qquad\qquad\quad\quad\!\!\!
        + \delta_{BA'} \, \B_{\lambda\ell\ell'}^{BAB'}
        + \delta_{BB'} \, \B_{\lambda\ell\ell'}^{BAA'}
    \Bigr\} \\
    &+ \nu^{AB}_{A'B'} \sum_{\lambda} Q_{\lambda\ell\ell'} \, \C_\lambda^{AB} \;.
\end{align}
Taking the expectation, we obtain the result that the covariance of the angular power spectra consists of true four-point terms, plus three- and two-point contributions from degenerate combinations with one and two repeated points, respectively.
These shot noise contributions to the covariance are fully determined by the bispectrum and power spectrum, respectively, and do not depend on any additional characteristics of the point process.

From expression~\eqref{eq:cov}, we are able to recover the usual `Gaussian covariance' approximation of shot noise: only on the diagonal $\ell' = \ell$ does the sum over~$\lambda$ extend to the monopole $\lambda = 0$ in both the two-point contribution,
\begin{equation}
    Q_{0\ell\ell} \, \C_0^{AB}
    = \frac{1}{2\ell + 1} \, \frac{N_A N_B - N_{AB}}{(4\pi)^2} \;,
\end{equation}
and the three-point contributions,
\begin{equation}
    R_{0\ell\ell} \, \B_{0\ell\ell}^{ABC}
    = \frac{1}{2\ell+1} \, \frac{N_A - \delta^{\rm K}_{AB} - \delta^{\rm K}_{AC}}{4\pi} \, \C^{BC}_\ell \;,
\end{equation}
where neither~$\C_0^{AB}$ nor~$\B_{0\ell\ell}^{ABC}$ contain true two- or three-point information.
These monopole terms correspond to the $(\text{shot noise})^2$ and $C_\ell \times \text{(shot noise)}$ terms of the Gaussian covariance approximation $(C_\ell + \text{shot noise})^2$.

\begin{figure}[t!]%
\centering%
\includegraphics[scale=0.8]{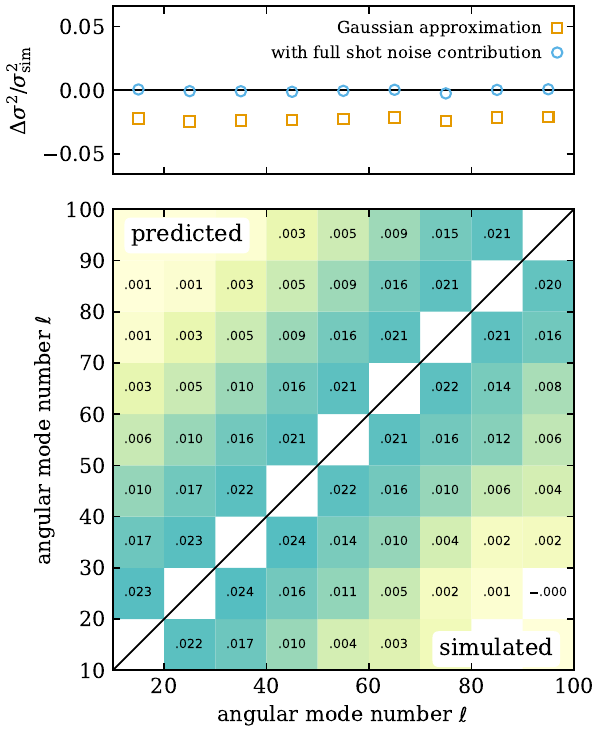}%
\caption{%
    Angular power spectrum covariance from 1\,000\,000 Gaussian full-sky simulations with Poisson sampling of 100 points.
    Results are binned as indicated.
    \emph{Top panel:}
    Relative error of the variance predicted by the Gaussian approximation and with the full shot noise contribution compared to the sample variance of the simulations.
    \emph{Bottom panel:}
    Correlation matrix predicted when including the full shot noise contribution (\emph{upper triangle}) and measured from the simulations (\emph{lower triangle}).
    The off-diagonal contributions are not captured by the Gaussian approximation.
}%
\label{fig:cov}%
\end{figure}

For practical computations, it is useful to note that the coefficients~\eqref{eq:q} are equivalent to those of a rescaled, and thus symmetrised, mixing matrix~$M_{\ell\ell'}^{AB}$ that is obtained by applying the usual mixing matrix formalism \citep[see, e.g.,][]{2025A&A...694A.141E} to~$\Ev{\C_\lambda^{AB}}$,
\begin{equation}
    \sum_{\lambda} Q_{\lambda\ell\ell'} \Ev{\C_\lambda^{AB}}
    = \frac{1}{2\ell'+1} \, M_{\ell\ell'}^{AB} \;.
\end{equation}
In particular, it follows that this shot noise contribution to a partial-sky covariance is a rescaled mixing matrix computed from an angular power spectrum that itself has the mixing matrix of the footprint applied.
Since tools for obtaining mixing matrices are widely available, this generally non-diagonal contribution to the covariance is readily computed in full.

To demonstrate the effect of the shot noise contribution to the covariance, we simulate 1\,000\,000 Gaussian random fields over the full sky.
Afterwards, we draw a mean number of 100 points over the full sky via Poisson sampling.
This low number is unrealistic, but ensures that the impact of shot noise is easy to detect even on large scales.
For the same reason, we employ a toy model for the simulated angular power spectrum,
\begin{equation}
    \Ev{C_\ell}
    = \begin{dcases}
        2 \times 10^{-4} \, (2\ell + 1) \, \E^{-\ell/10} &
        \text{if $\ell > 0$,} \\
        0 & \text{if $\ell = 0$,}
    \end{dcases}
\end{equation}
which results in significant off-diagonal terms in the two-point contribution in expression~\eqref{eq:cov}.
Since full-sky Gaussian random fields do not introduce three- or four-point correlations in the simulated point process, we expect to have an exact description of the covariance of our simulation.
This is shown in Fig.~\ref{fig:cov}, where we find excellent agreement between the predicted and simulated covariance. 

\begin{figure}[t!]%
\centering%
\includegraphics[scale=0.8]{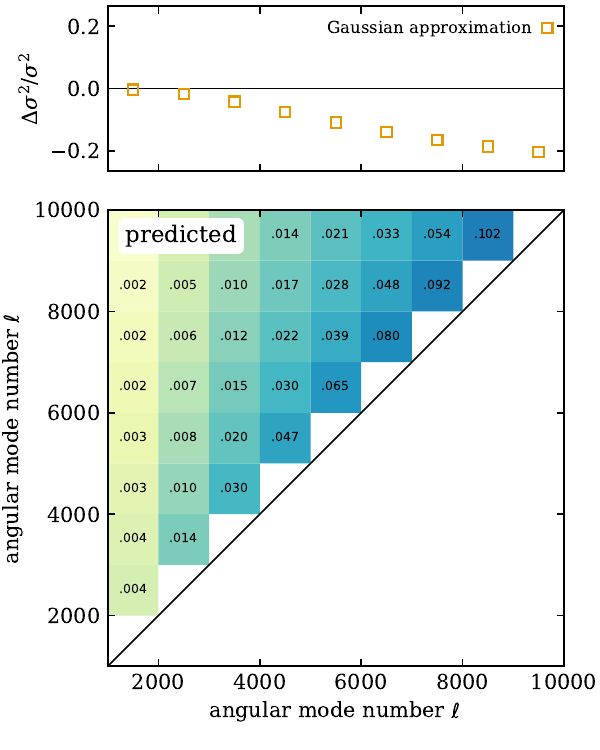}%
\caption{%
    Angular power spectrum covariance for a sample of 100 million points in a realistic setting.
    Results are binned as indicated.
    \emph{Top panel:}
    Relative error of the variance predicted by the Gaussian approximation compared to the variance with the full shot noise contribution.
    \emph{Bottom panel:}
    Correlation matrix predicted when including the full shot noise contribution.
}%
\label{fig:cov2}%
\end{figure}

Having thus validated our expression for the shot noise contribution to the covariance with a simple toy model, we now show that the impact of the shot noise terms beyond the Gaussian approximation is not necessarily small in a more realistic setting.
First, we obtain a theory angular power spectrum for galaxy clustering, including non-linear contributions to the matter power spectrum, using CAMB \citep{2000ApJ...538..473L, 2012JCAP...04..027H}.
We assume a representative but arbitrary $\Lambda$CDM cosmology with~$h=0.7, \Omega_m = 0.3$, $\Omega_b = 0.05$ and a Gaussian tomographic redshift bin centred on~$z = 0.5$ with standard deviation~$\sigma_z = 0.05$ and a linear galaxy bias~$b = 1.5$.
The resulting covariance for a Gaussian density field with 100 million points observed over the full sky is shown in Fig.~\ref{fig:cov2}.
When using a fairly coarse angular binning, as is standard practice, we find a significant deviation of the Gaussian approximation from the full expression on small angular scales where the shot noise bias is of the same order of magnitude as the signal.
There, we also find significant off-diagonal contributions to the covariance, which could be misidentified as an imprint of non-Gaussianity if not attributed to the additional shot noise terms in the covariance.
However, accurately demonstrating these effects with simulations is much more costly than for the simple toy example, and we leave a thorough exploration of the importance for Stage IV surveys to forthcoming work.

\section{Conclusions}
\label{sec:conclusions}

In this work, we have established some facts about shot noise bias that are perhaps not obvious at first sight.

Firstly, the shot noise contribution to the clustering power spectrum is independent of the distribution of the points and has a fixed and known value.
Any apparent deviation from that value points to a deviation of the two-point statistics from the expectation at small scales.
If the effect is localised below some scale~$\theta_*$ in real space, the power spectrum will reveal an effect that is imprinted on scales below~$\ell_* \sim \theta_*^{-1}$ in harmonic space.
This occurs even if the real-space scale~$\theta_*$ is too small to be observed directly.
An apparent non-Poissonian value of the shot noise could therefore be a valuable signal in itself, and inform us about effects such as assembly bias or halo exclusion.

Conversely, it may be possible to absorb unresolved small-scale effects into an `effective shot noise' contribution to the power spectra, as done in the effective field theory of large-scale structure \citep[e.g.,][]{2023JCAP...01..028C}.
But such an effective description is necessarily scale-dependent: extending the predictions to smaller scales will ultimately require reverting back to the Poissonian shot noise value, and resolving the small-scale effects into actual two-point statistics.
Overall, two point processes cannot differ only by shot noise: having more or less power on all (observable) scales means that there is a tangible difference in the clustering of points.

In simulations where the realised point process is the tracer of a discretised density field, it may be necessary to explicitly account for non-Poissonian distributions of the tracer.
This depends on whether the theoretical two-point statistics are those of the density or the point process: in the former case, the apparent non-Poissonian value of the shot noise contains real two-point information, as in the case of observations.
In the latter case, the two-point statistics of the density field must be modified to realise a point process with the intended two-point statistics.

In practical applications of power spectrum estimation, it is common to apply weights to galaxies to account for selection biases, correct for systematics, or improve the statistical performance of the estimator \citep{1994ApJ...426...23F, 2016MNRAS.463.2708P, 2017MNRAS.464.1168R}.
We have assumed unit weights in this work; in the case of non-unit weights, number counts that appear in our expressions ($N_A$, etc.) should be replaced by their effective (i.e., weighted) versions.
Weights introduce no new source of stochasticity, so the conclusions of this paper remain unchanged.

Similarly, we have not considered more complicated mass assignment schemes that may be employed by grid-based power spectrum estimators, and have instead focused on the grid-free catalogue-based estimator or the Nearest Grid Point (NGP) scheme.
As shown by \citet{2005ApJ...620..559J}, the shot noise bias has a simple form in the NGP scheme, but exact expressions can be derived for the commonly used higher-order schemes such as Cloud-in-Cell.
Issues to do with mass assignment do not impact the conclusions of this work, as they only amount to a deterministic modification of the harmonic-space form of the shot noise bias.

The shot noise contribution to the covariance of the observed angular power spectra can similarly be computed exactly for any point process: it only depends on the two-, three-, and four-point statistics, but no other characteristics.
This recovers the usual Gaussian covariance approximation as one term on the diagonal, but the full expression contains off-diagonal terms even in the case of Gaussian random fields and Poisson sampling \citep{1999MNRAS.308.1179M, 2006NewA...11..226C, 2018A&A...615A...1L, 2020A&A...634A..74L}.

Although we have derived results here for angular clustering, the derivation is identical for 3D clustering, if the spherical harmonics are replaced by plane waves, and the Legendre polynomials are replaced by the spherical Bessel function~$j_0(kr)$.
The only complication arises in the shot noise contribution to the covariance, where the usual extra care must be taken to account for the finite volume of the sample, as the shot noise contribution to the covariance vanishes over an infinite volume.
In summary, the results are conceptually identical for 3D clustering: the shot noise contributions to the power spectra and their covariance have known values that are independent of the nature of the point process beyond two-, three-, and four-point statistics.

\section*{Acknowledgements}

This note is the product of conversations about shot noise with Benjamin Joachimi, Isaac Tutusaus, Stefano Camera, David Alonso, Alkistis Pourtsidou, Benedict Bahr-Kalus, and Emiliano Sefusatti.
NT is grateful to David Alonso for proposing the grid-free measurement to show that the shot noise returns to the Poissonian value.
NT acknowledges support by the ERC-selected UKRI Frontier Research Grant EP/Y03015X/1.
NT further acknowledges support by the UK Space Agency through grants ST/W002574/1 and ST/X00208X/1.
AH is supported by a Royal Society University Research Fellowship.
For the purpose of open access, the authors have applied a Creative Commons Attribution (CC BY) licence to any Author Accepted Manuscript version arising from this submission.

\bibliographystyle{mnras}
\bibliography{main}

\begin{thebibliography}{}
\makeatletter
\relax
\def\mn@urlcharsother{\let\do\@makeother \do\$\do\&\do\#\do\^\do\_\do\%\do\~}
\def\mn@doi{\begingroup\mn@urlcharsother \@ifnextchar [ {\mn@doi@} {\mn@doi@[]}}
\def\mn@doi@[#1]#2{\def\@tempa{#1}\ifx\@tempa\@empty \href {http://dx.doi.org/#2} {doi:#2}\else \href {http://dx.doi.org/#2} {#1}\fi \endgroup}
\def\mn@eprint#1#2{\mn@eprint@#1:#2::\@nil}
\def\mn@eprint@arXiv#1{\href {http://arxiv.org/abs/#1} {{\tt arXiv:#1}}}
\def\mn@eprint@dblp#1{\href {http://dblp.uni-trier.de/rec/bibtex/#1.xml} {dblp:#1}}
\def\mn@eprint@#1:#2:#3:#4\@nil{\def\@tempa {#1}\def\@tempb {#2}\def\@tempc {#3}\ifx \@tempc \@empty \let \@tempc \@tempb \let \@tempb \@tempa \fi \ifx \@tempb \@empty \def\@tempb {arXiv}\fi \@ifundefined {mn@eprint@\@tempb}{\@tempb:\@tempc}{\expandafter \expandafter \csname mn@eprint@\@tempb\endcsname \expandafter{\@tempc}}}

\bibitem[\protect\citeauthoryear{{Baldauf}, {Seljak}, {Smith}, {Hamaus}  \& {Desjacques}}{{Baldauf} et~al.}{2013}]{2013PhRvD..88h3507B}
{Baldauf} T.,  {Seljak} U.,  {Smith} R.~E.,  {Hamaus} N.,   {Desjacques} V.,  2013, \mn@doi [\prd] {10.1103/PhysRevD.88.083507}, \href {https://ui.adsabs.harvard.edu/abs/2013PhRvD..88h3507B} {88, 083507}

\bibitem[\protect\citeauthoryear{{Baleato Lizancos} \& {White}}{{Baleato Lizancos} \& {White}}{2024}]{2024JCAP...05..010B}
{Baleato Lizancos} A.,  {White} M.,  2024, \mn@doi [JCAP] {10.1088/1475-7516/2024/05/010}, \href {https://ui.adsabs.harvard.edu/abs/2024JCAP...05..010B} {2024, 010}

\bibitem[\protect\citeauthoryear{{Carrilho}, {Moretti}  \& {Pourtsidou}}{{Carrilho} et~al.}{2023}]{2023JCAP...01..028C}
{Carrilho} P.,  {Moretti} C.,   {Pourtsidou} A.,  2023, \mn@doi [\jcap] {10.1088/1475-7516/2023/01/028}, \href {https://ui.adsabs.harvard.edu/abs/2023JCAP...01..028C} {2023, 028}

\bibitem[\protect\citeauthoryear{{Cohn}}{{Cohn}}{2006}]{2006NewA...11..226C}
{Cohn} J.~D.,  2006, \mn@doi [\na] {10.1016/j.newast.2005.08.002}, \href {https://ui.adsabs.harvard.edu/abs/2006NewA...11..226C} {11, 226}

\bibitem[\protect\citeauthoryear{{DESI Collaboration} et~al.,}{{DESI Collaboration} et~al.}{2025}]{2025arXiv250314738D}
{DESI Collaboration} et~al., 2025, \mn@doi [arXiv e-prints] {10.48550/arXiv.2503.14738}, \href {https://ui.adsabs.harvard.edu/abs/2025arXiv250314738D} {p. arXiv:2503.14738}

\bibitem[\protect\citeauthoryear{{Dark Energy Survey Collaboration} et~al.,}{{Dark Energy Survey Collaboration} et~al.}{2022}]{2022PhRvD.105b3520A}
{Dark Energy Survey Collaboration} et~al., 2022, \mn@doi [\prd] {10.1103/PhysRevD.105.023520}, \href {https://ui.adsabs.harvard.edu/abs/2022PhRvD.105b3520A} {105, 023520}

\bibitem[\protect\citeauthoryear{{Dvornik} et~al.,}{{Dvornik} et~al.}{2023}]{2023A&A...675A.189D}
{Dvornik} A.,  et~al., 2023, \mn@doi [\aap] {10.1051/0004-6361/202245158}, \href {https://ui.adsabs.harvard.edu/abs/2023A&A...675A.189D} {675, A189}

\bibitem[\protect\citeauthoryear{{Euclid Collaboration: Tessore} et~al.,}{{Euclid Collaboration: Tessore} et~al.}{2025}]{2025A&A...694A.141E}
{Euclid Collaboration: Tessore} N.,  et~al., 2025, \mn@doi [\aap] {10.1051/0004-6361/202452018}, \href {https://ui.adsabs.harvard.edu/abs/2025A&A...694A.141E} {694, A141}

\bibitem[\protect\citeauthoryear{{Feldman}, {Kaiser}  \& {Peacock}}{{Feldman} et~al.}{1994}]{1994ApJ...426...23F}
{Feldman} H.~A.,  {Kaiser} N.,   {Peacock} J.~A.,  1994, \mn@doi [\apj] {10.1086/174036}, \href {https://ui.adsabs.harvard.edu/abs/1994ApJ...426...23F} {426, 23}

\bibitem[\protect\citeauthoryear{{Ginzburg}, {Desjacques}  \& {Chan}}{{Ginzburg} et~al.}{2017}]{2017PhRvD..96h3528G}
{Ginzburg} D.,  {Desjacques} V.,   {Chan} K.~C.,  2017, \mn@doi [\prd] {10.1103/PhysRevD.96.083528}, \href {https://ui.adsabs.harvard.edu/abs/2017PhRvD..96h3528G} {96, 083528}

\bibitem[\protect\citeauthoryear{{Howlett}, {Lewis}, {Hall}  \& {Challinor}}{{Howlett} et~al.}{2012}]{2012JCAP...04..027H}
{Howlett} C.,  {Lewis} A.,  {Hall} A.,   {Challinor} A.,  2012, \mn@doi [\jcap] {10.1088/1475-7516/2012/04/027}, \href {https://ui.adsabs.harvard.edu/abs/2012JCAP...04..027H} {2012, 027}

\bibitem[\protect\citeauthoryear{{Jing}}{{Jing}}{2005}]{2005ApJ...620..559J}
{Jing} Y.~P.,  2005, \mn@doi [\apj] {10.1086/427087}, \href {https://ui.adsabs.harvard.edu/abs/2005ApJ...620..559J} {620, 559}

\bibitem[\protect\citeauthoryear{{Lacasa}}{{Lacasa}}{2018}]{2018A&A...615A...1L}
{Lacasa} F.,  2018, \mn@doi [\aap] {10.1051/0004-6361/201732343}, \href {https://ui.adsabs.harvard.edu/abs/2018A&A...615A...1L} {615, A1}

\bibitem[\protect\citeauthoryear{{Lacasa}}{{Lacasa}}{2020}]{2020A&A...634A..74L}
{Lacasa} F.,  2020, \mn@doi [\aap] {10.1051/0004-6361/201936683}, \href {https://ui.adsabs.harvard.edu/abs/2020A&A...634A..74L} {634, A74}

\bibitem[\protect\citeauthoryear{{Layzer}}{{Layzer}}{1956}]{1956AJ.....61..383L}
{Layzer} D.,  1956, \mn@doi [\aj] {10.1086/107366}, \href {https://ui.adsabs.harvard.edu/abs/1956AJ.....61..383L} {61, 383}

\bibitem[\protect\citeauthoryear{{Lewis}, {Challinor}  \& {Lasenby}}{{Lewis} et~al.}{2000}]{2000ApJ...538..473L}
{Lewis} A.,  {Challinor} A.,   {Lasenby} A.,  2000, \mn@doi [\apj] {10.1086/309179}, \href {https://ui.adsabs.harvard.edu/abs/2000ApJ...538..473L} {538, 473}

\bibitem[\protect\citeauthoryear{{Meiksin} \& {White}}{{Meiksin} \& {White}}{1999}]{1999MNRAS.308.1179M}
{Meiksin} A.,  {White} M.,  1999, \mn@doi [\mnras] {10.1046/j.1365-8711.1999.02825.x}, \href {https://ui.adsabs.harvard.edu/abs/1999MNRAS.308.1179M} {308, 1179}

\bibitem[\protect\citeauthoryear{{More} et~al.,}{{More} et~al.}{2023}]{2023PhRvD.108l3520M}
{More} S.,  et~al., 2023, \mn@doi [\prd] {10.1103/PhysRevD.108.123520}, \href {https://ui.adsabs.harvard.edu/abs/2023PhRvD.108l3520M} {108, 123520}

\bibitem[\protect\citeauthoryear{{Pashapour-Ahmadabadi}, {B{\"o}hme}, {Siewert}, {Schwarz}, {Hale}, {Heneka}, {Tiwari}  \& {Zheng}}{{Pashapour-Ahmadabadi} et~al.}{2025}]{2025A&A...698A.148P}
{Pashapour-Ahmadabadi} M.,  {B{\"o}hme} L.,  {Siewert} T.~M.,  {Schwarz} D.~J.,  {Hale} C.~L.,  {Heneka} C.,  {Tiwari} P.,   {Zheng} J.,  2025, \mn@doi [\aap] {10.1051/0004-6361/202452734}, \href {https://ui.adsabs.harvard.edu/abs/2025A&A...698A.148P} {698, A148}

\bibitem[\protect\citeauthoryear{{Pearson}, {Samushia}  \& {Gagrani}}{{Pearson} et~al.}{2016}]{2016MNRAS.463.2708P}
{Pearson} D.~W.,  {Samushia} L.,   {Gagrani} P.,  2016, \mn@doi [\mnras] {10.1093/mnras/stw2177}, \href {https://ui.adsabs.harvard.edu/abs/2016MNRAS.463.2708P} {463, 2708}

\bibitem[\protect\citeauthoryear{{Peebles}}{{Peebles}}{1973}]{1973ApJ...185..413P}
{Peebles} P.~J.~E.,  1973, \mn@doi [\apj] {10.1086/152431}, \href {https://ui.adsabs.harvard.edu/abs/1973ApJ...185..413P} {185, 413}

\bibitem[\protect\citeauthoryear{{Ross} et~al.,}{{Ross} et~al.}{2017}]{2017MNRAS.464.1168R}
{Ross} A.~J.,  et~al., 2017, \mn@doi [\mnras] {10.1093/mnras/stw2372}, \href {https://ui.adsabs.harvard.edu/abs/2017MNRAS.464.1168R} {464, 1168}

\bibitem[\protect\citeauthoryear{{Wolz}, {Alonso}  \& {Nicola}}{{Wolz} et~al.}{2025}]{2025JCAP...01..028W}
{Wolz} K.,  {Alonso} D.,   {Nicola} A.,  2025, \mn@doi [JCAP] {10.1088/1475-7516/2025/01/028}, \href {https://ui.adsabs.harvard.edu/abs/2025JCAP...01..028W} {2025, 028}

\makeatother
\end{thebibliography}

\end{document}